\documentclass{elsart}

\usepackage{amssymb}
\begin{document}

\begin{frontmatter}



\title{Proton structure function.\\ Soft and hard Pomerons\thanksref{talk}}
\thanks[talk]{Presented by E. Martynov}

\author[Desgr]{P. Desgrolard},
\author[Mart]{E. Martynov}
\address[Desgr]{Institut de Physique
Nucl\'eaire de Lyon, IN2P3-CNRS et Universit\'e Claude Bernard, 43
boulevard du 11 novembre 1918, F-69622 Villeurbanne Cedex, France}
\ead{desgrolard@inpnl.in2p3.fr({\rm P. Desgrolard})}
\address[Mart]{ Bogoliubov Institute for Theoretical Physics, National
Academy of Sciences of Ukraine, 03143 Kiev-143, Metrologicheskaja 14b,
Ukraine}
\ead{martynov@bitp.kiev.ua}
\begin{abstract}
Regge models for proton structure function with and without a hard
Pomeron contribution are compared with all available data in the region
$W>3$ GeV, $Q^{2}\leq 3000$ GeV$^{2}$ and $x<0.75$. It is shown that
the data do not support a hard Pomeron term in
$\gamma^{*} p$ amplitude. Moreover, the data support the idea that
the soft Pomeron, either is a double pole with $\alpha_{P}(0)=1$ in the
angular momentum $j$-plane, or is a simple pole with
$\alpha_{P}(0)=1+\epsilon$ where $\epsilon \ll 1$.
\end{abstract}


\end{frontmatter}


The Regge approach to Deep Inelastic Scattering (DIS) is based on
several facts and assumptions such as the similarity of photon and
hadrons at high energies, the one-photon approximation, the optical
theorem, the relation between elastic forward $\gamma p$ amplitude and
structure functions (SF). The most important assumptions for the Regge
description of DIS are the analyticity of $\gamma p$ amplitude and the
properties of a dominating contribution, namely, of a Pomeron. One
should have in mind that the very reasonable assumption about
analyticity is only an assumption. The unitarity restrictions on the
amplitude and on the related observable quantities are not proved in
the case of DIS while they are often supposed to be valid, at least at
$Q^{2}\sim 0$ and $x\to 0$. The universality of the Pomeron and of the
Reggeons are crucial ingredients in Regge approach being especially
important if it is applied to DIS. This means that the Reggeons are the
same in pure hadronic and in lepton-hadronic processes, their
trajectories can not depend on the individual properties of interacting
particles.

The starting point in Regge approach to DIS is a choice of a model for
the Pomeron. There are two main possibilities used in hadron physics as
well as in DIS. Both of them take into account the fact that the total
cross-sections are rising with energy. The first one assumes that the
input Pomeron behaves as a simple $j$-pole and that its trajectory has
an intercept $\alpha_{P}(0)=1+\Delta>1$. Such a Pomeron leads to
$\sigma_{tot}\propto s^{\Delta}$, when $s\to \infty$, violating the
unitarity restriction $\sigma_{tot}\leq C\ell n^{2}(s/m^{2})$. In order
to restore unitarity, the Pomeron must be unitarized. An other
possibility to describe the growing total cross-sections is to assume
that the Pomeron is a more hard singularity than a simple pole. For
instance, it can be a double pole (then, $\sigma_{tot}\propto \ell n
s$) or a triple pole (then, $\sigma_{tot}\propto \ell n^{2}s$). Such a
Pomeron does not violate the Froissart-Martin bound. Both types of
Pomeron models have been developed and compared to the available data
of the total cross-sections of nucleon-nucleon, meson-nucleon,
photon-proton and photon-photon interactions
\cite{DGLM,CEKLT,DGMP-ed,Nicol,Kang,CEGKKLNT}. The main lessons of
these investigations are the following.\\
 1. A better description of the data on total cross-sections and
$\rho$-ratios of real to imaginary part of the forward scattering
amplitudes is achieved in Pomeron models of the second type. Moreover
the Dipole Pomeron (or double $j$-pole with $\alpha_{P}(0)=1$) model
gives the minimal
$\chi^{2}$ \cite{DGLM,CEKLT,DGMP-ed,CEGKKLNT}.\\
 2. If a simple pole Pomeron contribution, $C(-is/s_{0})^{\Delta}$ is
 generalized \cite{DGLM,DGMP-ed} into
$C_{0}+C_{1}(-is/s_{0})^{\Delta}$, then fitting the data leads to
$\Delta \sim 0.001$ ($s_0$ is fixed at 1 GeV$^2$). Practically this
"supercritical" Pomeron reduces to the Dipole Pomeron because
$s^{\Delta}\approx 1+\Delta \ell ns$ at $\Delta \ll 1$
\cite{DGLM,DGMP-ed}.

Having in mind the above circumstances and the quite good description of
the data on proton SF within the Soft Dipole Pomeron model
\cite{DLM-SDP} we verify the validity of the above property (2)
in DIS when $Q^{2}\neq 0$. At the same time using a common data
set we compare models for SF with and without a hard Pomeron
contribution. This last component was introduced \cite{DL2pom} to
describe a steep rise of SF when $Q^{2}$ is large and $x\to 0$. In our
opinion however, adding a hard Pomeron component contradicts the
idea of Pomeron universality because it is not observed in pure hadron
processes \cite{Nicol,Kang,CEGKKLNT} and in $\gamma p$ interaction at
$Q^{2}=0$. Finally, we suggest a new model for the proton SF that does not
violate unitarity restrictions for cross-section but at high $Q^{2}$
mimics the contribution of a hard Pomeron with $\Delta \approx 0.4$.
\section{Models for proton SF at $x\ll 1$}\label{Models}
{\bf Donnachie-Landshoff Soft+Hard Pomeron ({\rm SHP}) model}\\
The model is constructed as sum of two Pomeron components,
a hard one $F_{hard}$ and a soft one $F_{soft}$. The contribution of an
$f(a_{2})$-Reggeon is necessary to describe the structure function and
the $\gamma p$ total cross-section at low energies.
\begin{equation}\label{eq:F2-SHP}
F_{2}(x,Q^{2})=F_{hard}+F_{soft}+F_{f}\ .
\end{equation}
We have considered the version published by Donnachie and Landshoff
(D-L) in \cite{DL2pom}.
{\small
\begin{equation}\label{eq:hard-SHP} F_{hard}=C_{h}\left(
\frac{Q^{2}}{Q^{2}+Q^{2}_{h}}\right)^{1+\epsilon_{h}}
\left(1+\frac{Q^{2}}{Q^{2}_{h}}\right)^{\frac{1}{2}\epsilon_{h}}
\left(\frac{1}{x}\right)^{\epsilon_{h}},
\end{equation}
\begin{equation}\label{eq:soft-SHP}
F_{soft}=C_{s}\left( \frac{Q^{2}}{Q^{2}+Q^{2}_{s}}\right)^{1+\epsilon_{s}}
\left(\frac{1}{x}\right)^{\epsilon_{s}},
\end{equation}
\begin{equation}\label{eq:f-SHP}
F_{f}=C_{f}\left( \frac{Q^{2}}{Q^{2}+Q^{2}_{f}}\right)^{\alpha_{f}}
\left(\frac{1}{x}\right)^{\alpha_{f}-1}.
\end{equation}
}
One can easily obtain from (\ref{eq:F2-SHP})-(\ref{eq:f-SHP}) the
expression for $\gamma p$ total cross-section
\begin{equation}\label{stot-SHP}
\sigma_{T}(W)= \frac{4\pi^{2}\alpha}{Q^{2}}F_{2}(x,Q^{2})\bigg
|_{Q^{2}=0}= 4\pi^{2}\alpha\sum\limits_{i=h,s,f}
\frac{C_{i}}{(Q_{i}^{2})^{1+\epsilon_{i}}}(W^{2}-m_{p}^{2})^{\epsilon_{i}}.
\end{equation}
where $\epsilon_{f}=\alpha_{f}-1.$

{\bf Soft Dipole Pomeron ({\rm SDP}) model}\\
This model constructed in \cite{DLM-SDP} makes use
of the optical theorem and of the relation between elastic forward
$\gamma^{*}p$ amplitude and structure function: {\small
\begin{equation}\label{stot-F2}
 \sigma_{T}^{\gamma^*p}=8\pi\Im
m A(W^2,Q^2;t=0)=\frac{4\pi^2\alpha}{Q^2(1-x)}\left
(1+\frac{4m_p^2x^2}{Q^2}\right )F_2(x,Q^2) ;
\end{equation}
\begin{equation}\label{ampl-SDP}
 A(W^2,t=0;Q^2)=P_{0}+P_{1}+f,
\end{equation}
where
\begin{equation}\label{f-SDP}
f=iG_f(Q^2)(-iW^2/m_p^2)^{\alpha_f-1}(1-x)^{B_f(Q^{2})}
\end{equation}
}{\small
\begin{equation}
P_0=iG_0(Q^2) (1-x)^{B_0(Q^2)},\quad P_1=iG_1(Q^2)\ell
n\left(-i\frac{W^{2}}{m_{p}^{2}}\right)(1-x)^{B_1(Q^2)}\nonumber
\end{equation}
{\small
\begin{equation}\label{G,D-SDP}
G_i(Q^2)=\frac{C_i}{\left(1\!+\!Q^2/Q_{i}^2 \right)^{D_i(Q^2)}},
\quad D_i(Q^2)=d_{i\infty}+\frac{d_{i0}-d_{i\infty}}{1\!+\!Q^2/ Q_{id}^2},
\end{equation}
}
{\small
\begin{equation}\label{B-Q}
B_i(Q^2)=b_{i\infty}+\frac{b_{i0}-b_{i\infty}}{1+Q^2/Q^2_{ib}}.
\end{equation}
}

We would like to emphasize that here the Pomeron is a double $j$-pole,
the trajectory of which being $Q^{2}$-independent with an intercept
$\alpha_{P}(0)=1.$ The intercept of $f$-reggeon was fixed at 0.785 as
determined \cite{DGMP-ed} from the fit to total cross-sections.

{\bf Modified two-Pomeron ({\rm Mod2P}) model}\\
Modifying the SHP model we aimed to verify the phenomenon
mentioned above for the total cross-sections of hadrons and real photons
when adding a Pomeron component (simple $j$-pole with an intercept one)
leads to a very small value of $\Delta =\alpha_{P}(0)-1$. So we construct
the model as a sum of two Pomeron terms (the first one has a unit intercept
while the other one has $\alpha_{P}(0)>1$). Comparing with SHP model
the residues $C_{i}(Q^{2})$ are chosen dimensionless and modified as
shown below.

{\small
\begin{equation}
F_{2}(x,Q^{2})=F_{0}+F_{s}+F_{f},
\end{equation}
\begin{equation}
F_{0}=\frac{C_{0}Q^{2}_{0}}{\Delta }\left(
\frac{Q^{2}}{Q^{2}+Q^{2}_{0}}\right)
\left(1+\frac{Q^{2}}{Q^{2}_{0\,1}}\right)^{d_{0}},
\end{equation}
\begin{equation}
F_{s}=\!\frac{C_{s}Q^{2}_{s}}{\Delta (
m^{2}/Q^{2}_{s})^{\Delta}}\!\left(
\frac{Q^{2}}{Q^{2}\!+\!Q^{2}_{s}}\right)^{1+\Delta}\!
\left(\frac{1}{x}\right)^{\Delta}\!
\left(1\!+\!\frac{Q^{2}}{Q^{2}_{s\,1}}\right)^{d_{s}},
\end{equation}
\begin{equation}
\displaystyle
F_{f}=\frac{C_{f}Q^{2}_{f}}{(m^{2}/Q^{2}_{f})^{\alpha_{f}-1}}\left(
\frac{Q^{2}}{Q^{2}+Q^{2}_{f}}\right)^{\alpha_{f}}
\left(\frac{1}{x}\right)^{\alpha_{f}-1} \left(1+\frac{Q^{2}}{Q
_{f1}^{2}}\right)^{d_{f}}.
\end{equation}
\begin{equation}
\sigma_{T}(W)=4\pi^{2}\alpha \left \{ \frac{C_{0}}{\Delta}+
\frac{C_{s}}{\Delta}\left(\frac{W^{2}}{m^{2}}-1\right)^{\Delta}\right .
+ C_{f}\left(\frac{W^{2}}{m^{2}}-1\right)^{\alpha_{f}-1} \Bigg \}.
\end{equation}
}

Parameter $\Delta=\alpha_{P}(0)-1$ and intercept of $f$-reggeon are fixed
from the fit to all total cross-sections \cite{DGMP-ed}, namely
$\Delta =0.001013$ anf ${\alpha_{f}(0)=0.7895}$.

{\bf Generalized Logarithmic Pomeron ({\rm GLP}) model}\\
We have found in \cite{DLM-bx} a shortcoming of the SDP model, relative
to the logarithmic derivative $B_{x}=\partial \ell
nF_{2}(x,Q^{2})/\partial \ell n(1/x)$ at large $Q^{2}$ and small $x$.
Namely, in spite of a good $\chi^{2}$ in fitting the SF ($F_2$), theoretical
curves for $B_{x}$ are systematically slightly lower than the data
extracted from $F_{2}$. In our opinion, one reason might
be an insufficiently fast growth of $F_{2}$ with $x$ at large $Q^{2}$.
and small $x$ (the SDP model leads to a logarithmic behaviour in $1/x$)
On the other side, essentially a faster growth of $F_{2}$ (and
consequently of $B_{x}$) is, from a phenomenological point of view, a
good feature of the D-L model. However as stressed above, the hard Pomeron
component of this model contradicts the Pomeron universality and the
data on total cross-sections. Moreover it leads (see below) to a worse
$\chi^{2}$ than SDP does.

Thus, we have tried to construct a model that incorporates a slow rise
of $\sigma_{tot}^{\gamma p}(W)$ and simultaneously a fast rise of
$F_{2}(x,Q^2)$ at large $Q^{2}$ and small $x$. We propose below a model
intended to link these desirable properties

{\small
\begin{equation}
F_{2}(x,Q^{2})=F_{0}+F_{s}+F_{f},
\end{equation}
\begin{equation}
F_{f}=C_{f}\frac{Q^{2}}{(1\!+\!Q^{2}/Q^{2}_{f})^{d_{f}}}
\left(\frac{Q^{2}}{xm^{2}}\right)^{\alpha_{f}-1}\!\!(1-x)^{B_{f}(Q^{2})},
\end{equation}
\begin{equation}
F_{0}=C_{0}\frac{Q^{2}}{\left(1+Q^{2}/Q^{2}_{0}\right)^{d_{0}}}
(1-x)^{B_{0}(Q^{2})},
\end{equation}
\begin{equation}\label{F-sGLP}
F_{s}=C_{s}\frac{Q^{2}}{(1+Q^{2}/Q^{2}_{s })^{d_{s}}}L(W^{2},Q^{2})
(1-x)^{B_{s}(Q^{2})}
\end{equation}
}where $B_{i}(Q^{2}), \quad i=0,s,f $ are defined in accordance with
(\ref{B-Q}) and
{\small
\begin{equation}\label{LogGLP}
L(W^2,Q^{2})=\ell n
\left[1+\frac{a}{(1+Q^{2}/Q^{2}_{s\,l})^{d_{sl}}}\left(
\frac{Q^{2}}{xm^{2}}\right)^{\epsilon}\,\right ]\ .
\end{equation}
}
At $Q^{2}=0$, we have $L(W^{2},0)\approx \epsilon\ell
n(W^{2}/m_p^{2})$ when $W^{2}/m_p^{2}\gg 1$. Thus,
$\sigma_{tot}^{\gamma p}(W)\propto \ell n W^{2}$ at $W^{2}\gg
m_p^{2}$. For $Q^{2}\neq 0$ the logarithmic factor (\ref{LogGLP}) has
the following behavior
{\small
\begin{equation} L(W^{2},Q^{2})\approx \left \{
\begin{array}{ll}
\epsilon \,\ell n \left (\frac{W^{2}+Q^{2}}{m^{2}}-1\right),
\quad {\rm at}\quad Q^{2}/Q_{sl}^{2}\lesssim 1\\
\frac{a}{(1+Q^{2}/Q_{sl}^{2})^{d_{sl}}} \left
(\frac{W^{2}+Q^{2}}{m^{2}}-1\right )^{\epsilon}, \quad Q^{2}/Q_{sl}^{2}\gg
1
\end{array}
\right .
\end{equation}
} Thus the term (\ref{F-sGLP}) simulates a Pomeron contribution with
intercept $\alpha_{P}(0)=1+\epsilon$. We should emphasize that, in
spite of its appearance, the GLP model cannot be treated as a model
with a hard Pomeron, even when $\epsilon$ issued from the fit
($\epsilon \approx 0.32$ \cite{DM}) is not small. For the $f$-reggeon
intercept the fixed value 0.785 (as in SDP model) was used.

\section{Comparison of the models with data}
Fitting parameters of the models, we used all available data on
$\sigma_{tot}^{\gamma p}$ and $F_2$ in the region $W\geq 3$ GeV,
$Q^{2}\leq 3000$ GeV$^{2}$ and $x\leq 0.75$. All details concerning the
choice of data, corresponding references as well as the values of
parameters for each model and figures illustrating the good agreement
of the predictions with experimental points can be found in \cite{DM}.
Here we present only the main results and conclusions.

The values of $\chi^{2}/dof$ ($dof$ means degree of freedom $=$ number
of experimental points $-$ number of free parameters) representing a
good indicator or a confidence level in the models are given in the
Table for three cases. The two first fits (A and B) were made in the
region of small $x$ while the third fit (C) was performed for $x\leq
0.75$. When the models SDP and GLP were fitted at small $x$ the factors
$(1-x)^{B_{i}(Q^{2})}$ were set $\equiv 1$ in the expressions for SF.
Intercepts are chosen as it is explained above.

Thus from a comparison of the models under interest with the data on
structure functions at small $x$ we can made the following
conclusion:\\
all structure function data at $Q^{2}\leq 3000$ GeV$^{2}$ and small $x$
are described with a high quality {\it \bf w i t h o u t \hskip 0.3cm a
\hskip 0.3cm h a r d \hskip 0.3cm p o m e r o n}. Moreover, these data
support the idea that the soft Pomeron, either is a double pole with
$\alpha_{P}(0)=1$ in the angular momentum $j$-plane or is a simple pole
having intercept $\alpha_{P}(0)=1+\epsilon$ with a very small $\epsilon
$.

\smallskip

\begin{table}[h,t]\label{Table}
\caption{The results of the fits of 4 Regge models to the
small-$x$ (in two regions of energy) and large-$x$ SF data .}
\begin{center}
{\renewcommand{\arraystretch}{1.2}%
\begin{tabular}{|c||c|c|c|}
\hline
   &\multicolumn{3}{c|}{$\chi^{2}/dof$}\\
\cline{2-4}
   &\multicolumn{2}{c|}{$x\leq 0.07$}&{$x\leq 0.75$}\\
\cline{2-4}
Model                    &   Fit A   &   Fit B    & Fit C\\
                       & $W\geq 6$ GeV & $W\geq 3$ GeV& $W\geq 3$ GeV\\
\hline
SHP (with hard Pomeron)      &    1.375  &    1.450  & -\\
\hline
SDP (no hard Pomeron)        &    0.945  &    0.976  & 1.073\\
\hline
Mod2P (no hard Pomeron)      &    0.996  &    1.023  &-\\
\hline
GLP  (no hard Pomeron)       &    0.941  &    0.9685  & 1.064\\
\hline
\end{tabular}
}
\end{center}
\end{table}

\end{document}